\documentstyle[psfig]{l-aa} 
\def\MS{`Main Structure'{}}
\def\CS{`Foreground Structure'{}}
\def\a85{ABCG~85{}}
\def\CSS{`Central Substructure'{}}
\def\h50{h$_{50}^{-1}${}}
\def\large{`Large scale image'{}}
\thesaurus{11.03.1, 11.03.4,12.04.1, 13.25.2 }

\begin{document}
\title{The rich cluster of galaxies ABCG~85. I. X-ray analysis.
\thanks{Based on ROSAT Archive data.}}

\author {
  V.~Pislar\inst{1}
\and
  F.~Durret \inst{1,2}
\and
  D.~Gerbal \inst{1,2}  
\and
  G.~B.~Lima Neto \inst{1,3,4}
\and
 E.~Slezak \inst{5}
}
\offprints{ V.~Pislar, pislar@iap.fr }
\institute{
	Institut d'Astrophysique de Paris, CNRS, Universit\'e Pierre et 
Marie Curie, 98bis Bd Arago, F-75014 Paris, France 
\and 
	DAEC, Observatoire de Paris, Universit\'e Paris VII, CNRS (UA 173), 
F-92195 Meudon Cedex, France 
\and
	Observatoire de Lyon, Place Charles Andr\'e, F-69561 St Genis Laval
Cedex, France
\and
	Universit\"at Potsdam, c/o Astrophysikalisches Institut Potsdam,
	An der Sternwarte, 16, D-14882 Potsdam, Germany
\and
    	Observatoire de la C\^ote d'Azur, B.P. 229, F-06304 Nice Cedex 4, 
France 
}
\date{Received,  ; accepted,}

\maketitle

\begin{abstract}
We present an X-ray analysis of the rich cluster \a85 based on ROSAT 
PSPC data.  By applying an improved wavelet analysis, we show that our view 
of this cluster is notably changed from what was previously believed (a main 
region and a south blob). The 
main emission comes from the central part of the main body of the 
cluster on which is superimposed that of a foreground group of galaxies. The 
foreground group and the main cluster are separated (if redshifts are
cosmological) by 46~\h50 Mpc.  The southern blob is clearly not a group: 
it is resolved into X-ray emitting galaxies (in particular the second 
more luminous galaxy of the main cluster).  Several X-ray features are 
identified with bright galaxies.  We performed a spectral analysis and 
derived the temperature (T), metallicity (Z) and hydrogen column 
density (N$_{\rm H}$). The global quantities are: T=4~keV (in 
agreement with the velocity dispersion of 760~km/s) and Z=0.2Z$_\odot$.  We cannot derive accurate 
gradients for these quantities with our data, but there is strong 
evidence that the temperature is lower ($\sim 2.8$~keV) and the 
metallicity much higher (Z $\sim 0.8$Z$_\odot$) in the very centre (within about 
50~\h50 kpc).  We present a pixel by pixel method to model the physical 
properties of the X-ray gas and derive its density distribution.  We 
apply classical methods to estimate the dynamical, gas and stellar 
masses, as well as the cooling time and cooling flow characteristics.  At the 
limiting radius of the image (1.4~\h50 Mpc), we find M$_{\rm Dyn}\sim 
(2.1-2.9)10^{14}$~\h50 M$_{\odot}$, M$_{\rm gas}$/M$_{\rm Dyn}\sim 0.18 
h_{50}^{-1.5}$.  The stellar mass is $6.7\ 10^{12}$M$_{\odot}$, giving 
a mass to light ratio of M/L$_{\rm V}\sim 300$.  The cooling time is 
estimated for different models, leading to a cooling radius of 30-80~kpc
depending on the adopted cluster age; the mass deposit rate is 
20-70~M$_{\odot}$/yr, lower than previous 
determinations.  These results are discussed (cooling flow paradigm 
in relation with high Z, `baryonic crisis' etc.) in connection with 
current ideas on dynamical and evolutionary properties of clusters.

\end{abstract}

\section{Introduction}
The study of clusters and groups of galaxies, which contain 
a noticeable fraction of the galaxies in the Universe, is particularly 
important since they are astrophysical objects with the following 
properties: 
\begin{itemize}
\item Characteristical times of processes at work are 
in general long, with the result that clusters are witnesses of the 
current formation of structures in the Universe; their analysis may therefore
provide constraints on cosmological models.
\item They contain the main 
components of the Universe: baryonic matter (under stellar form and X-ray 
emitting plasma) and unseen matter, reaching a very high density.
\end{itemize}
This explains why processes in which these components interact, observed 
under the generic term of environment phenomena, are numerous: 
morphological segregation of galaxies, hydrogen deficiencies, steep 
slope of the luminosity function, etc.

Ideally, a thorough study of a cluster should be done by using simultaneously 
all types of data. We have therefore undertaken a complete analysis of 
several clusters, both in X-rays and through optical imaging and spectroscopy.
The X-ray emitting gas is a good tracer of the structure and morphology of
clusters, in particular of their gravitational 
potential, and as such has been extensively studied, using data from 
satellites with increasing sensitivities, as well as better spatial and energy 
resolutions. However, processing methods are so different in the X-rays and
optical that we will present our analysis of the cluster \a85 in two papers, 
the present one based on X-ray data, and the second one on optical data.

\a85 is a rich cluster of richness class 1 (Abell 1958), classified
as Bautz-Morgan type I (Leir \& van den Bergh 1977) and of RS cD type
(Struble \& Rood 1987). Its optical redshift is 0.0555, dominated by a 
cD (Colless, private communication). A spectrum of this galaxy kindly provided
by Colless prior to publication does not show characteristics of an active
galactic nucleus, in disagreement with Edge et al. (1992). Previous studies 
have claimed the existence in the cluster of a double structure in X-rays, 
including a main structure and a south blob (Forman \& Jones 1982, Gerbal et 
al. 1992, hereafter GDLL), and of a central cooling flow (Stewart et al. 1984, 
Prestwich et al. 1995). From Einstein data, GDLL  have derived the main 
physical parameters of the X-ray emitting gas, and have found in particular a 
very small value for the core radius of the X-ray gas 
($\sim 40\ h_{50}^{-1}$~kpc); they have shown the temperature to be roughly 
constant throughout the cluster, and the dark matter to be more concentrated 
towards the center of the cluster than the X-ray gas. 

The high spatial resolution observation of \a85 with the ROSAT HRI by 
Prestwich et al. (1995) has shown the possible existence of small bright 
emitting features (of sizes 5-10'', or $8-16h_{50}^{-1}$~kpc) which cannot 
be identified with single galaxies and are not likely to be foreground or 
background sources. They argue that these blobs could be either cooler and 
denser than the surrounding X-ray gas, or hotter and denser gas compressed by 
magnetic fields.

The energy range of ROSAT (E$<$2.5~keV) is not best suited for the observation 
of rich clusters, since in these objects the X-ray gas has a temperature 
larger than 2.5~keV, and therefore spectra corresponding to different gas 
temperatures look very similar in the ROSAT energy range. However, this 
satellite has allowed to map clusters with unprecedented spatial resolution, 
and even to derive information on possible spatial variations of the 
temperature (Briel \& Henry 1994, Henry \& Briel 1995), therefore showing that 
it can give useful information on clusters in spite of its energy range. 
>From ROSAT PSPC data, Kneer et al. (1996) have recently suggested that \a85 is 
not a fully relaxed system; they also claimed that the temperature of the 
X-ray gas is smaller in the center and in the south subclump than in the 
overall cluster. They identified this south blob with a loose group of 
galaxies that could be falling towards the center of the cluster. We will 
discuss this possibility in section~3.

We present here the results of our X-ray analysis of \a85 based on ROSAT PSPC 
data. Spatial information is derived using a multiscale (wavelet) 
analysis, in order to extract substructure at different spatial scales; 
this analysis has been improved since that presented in Slezak et al. (1994).
We present a pixel by pixel modelling of the X-ray gas and derive its 
physical properties (temperature, density, mass), adapted from a previous
method used to analyze Einstein data. Using the hydrostatic equilibrium hypothesis, 
we derive the cluster binding mass, the cooling time and cooling flow
parameters, as well as the predicted temperature decrease of the microwave
background by the Sunyaev-Zel'dovich effect. The optical imaging and 
spectroscopic catalogues, as well as their physical interpretation will be 
presented in forthcoming papers by Durret et al. and Slezak et al. 

We will show that the wavelet analysis of the X-ray
image, combined with a detailed physical analysis can strongly change 
our general vision of \a85. In section~2, we present the data and describe the
bases and behaviours of the three main techniques of analysis used here:
wavelet analysis, spectral analysis and modelling of the X-ray gas. Each
of these methods have led to various results described in the three
following sections. A critical review and discussion of these results is
presented in section~6, leading us to draw a synthetic picture of \a85,
and to derive consequences on the formation and evolution of this cluster.

\section{The data and methods of analysis}\label{method} 
\subsection{The data }\label{data}
The \a85 field was observed by the ROSAT PSPC at three separate times, 
Dec.  20, 1991, Jun.  11, 1992 (Sequence number wg800174P, PI R. Schwarz, 
hereafter the German image) and Jul.  1-2, 1992 (us800250P, 
PI C.  Jones, hereafter the US image).  The center of the pointed 
observations is 00h~41mn~50.4s, $-09^\circ$~17'24'' for the first two 
images and 0h~41mn~50.4s, $-09^\circ$~17'00'' for the third (equinox 
J2000.0).  The exposure times are 15949, 5709 and 10240 seconds 
respectively.  In all of these observations, the PSPC-B was used and 
the gain was at low state.  
Pixels were rebinned to a size of 15'', corresponding to 
$24.0~h_{50}^{-1}$~kpc at the cluster distance (taking
H$_0$=50~km~s$^{-1}$~Mpc$^{-1}$). We have defined the image limiting radius
R$_{\rm L}$ (15') as the radius where the number of photons reaches the image
background (i.e. about 1~photon/pixel). The surface brightness analysis was 
performed in the energy range 0.5-2.0~keV. All the analyses were
performed on the sum of the two images to minimize the 
uncertainties on the parameters, except for the wavelet analysis.

\subsection{Wavelet analysis}\label{ondelettes}
The raw X-ray images of the ABCG~85 cluster of galaxies have been 
processed using a multiscale vision model in order to remove the high 
frequency noise while keeping the small scale details linked to 
genuine objects.  The technique relies on an iterative 
object-by-object restoration procedure based on the hierarchical 
connection trees which define the objects in a thresholded wavelet 
space.  The basics of the algorithms involved are described in Slezak 
et al.  (1994) and Biviano et~al.  (1996), while a full description 
can be found in Ru\'e \& Bijaoui (1996).  Among the key points are the 
use of the wavelet transform to locate structures at various scales 
from a field labelling and the selection of objects by looking at the 
maximum wavelet coefficient values inside connected regions belonging to the 
same interscale connectivity graph.  Superimposed objects are then detected 
(at a 3$\sigma$ level) according to their typical scale, and accurately 
restored from the information at both this scale and the next smaller 
one, providing that their wavelet amplitudes are locally high enough 
to generate distinct peaks in a maximum wavelet value vs.  scale plot.  
The analysis results in a set of individual objects with different 
typical sizes, which are placed at their respective positions as they 
are restored, thereby generating the processed image.  Taking benefit 
of this additive process, a given set of objects is easily removed 
from the solution, either by not taking them into account initially, 
or by subtracting them after the generation of the processed image.  
Note that only resolved objects which are properly restored can be 
subtracted without inducing artifacts such as the ``hole'' described 
in section~3, or even negative values.

The linearity of the wavelet transform may however lead to 
misdetections for some peculiar object configurations.  A bright 
enough feature can indeed generate at all scales wavelet coefficients 
much higher than those computed for an underlying larger and fainter 
structure.  Although this large component has to be considered as a 
distinct object, it will remain undetected when the scheme summarized 
above is applied.  The X-ray image of A85 exhibits such a case since a 
very bright and unresolved peak is superimposed onto a diffuse 
component linked to the overall cluster.  In order to detect and 
restore such extended background features, we decided to overcome this 
limitation by applying our model twice.  At the end of the first step 
the restored objects are subtracted from the raw image and the 
resulting frame is searched for the objects which were missed because 
of the intensity of the now removed bright ones.  If an object is 
detected, it is restored and added to the previous set of objects, 
thereby building the final image.  Note that the spectral information 
has been lost in this final image.

\subsection{Spectral analysis}\label{spectre}
The EXSAS package (Zimmerman et al. 1994)
allows us to determine spectral properties of the X-ray emitting gas, for the 
cluster as a whole or for various parts of the 
cluster, with the requirement of having enough counts to perform an 
analysis.  The use of the package is well described in the EXSAS 
manual, thus we will only discuss some behaviours of interest for our 
purpose.  The model we use to fit the X-ray spectra is a Raymond-Smith 
plasma model (Raymond \& Smith 1977) which assumes bremsstrahlung 
emission and neutral hydrogen absorption, and is characterized by five 
parameters: the temperature, metal abundance, column density of 
neutral hydrogen along the line of sight $N_{H}$, redshift, and a 
normalization factor.  We have fixed the redshift to that determined 
optically (z=0.0555), leaving four free parameters.  The background was 
taken in a circle of radius 6' and with its center at 31' from the 
center of the cluster, where there is no visible point source.  The 
fit was done by a standard $\chi ^2$ minimization. We have corrected the 
images for vignetting and dead time, and rebinned the spectrum to obtain a 
signal to noise ratio of 5.

We have searched for correlations between parameters in the fitting process, 
using the whole cluster image (excluding the substructure south of the main 
body, see section~\ref{resultats}).  We find: 
\begin{figure}
        \centerline{\psfig{file=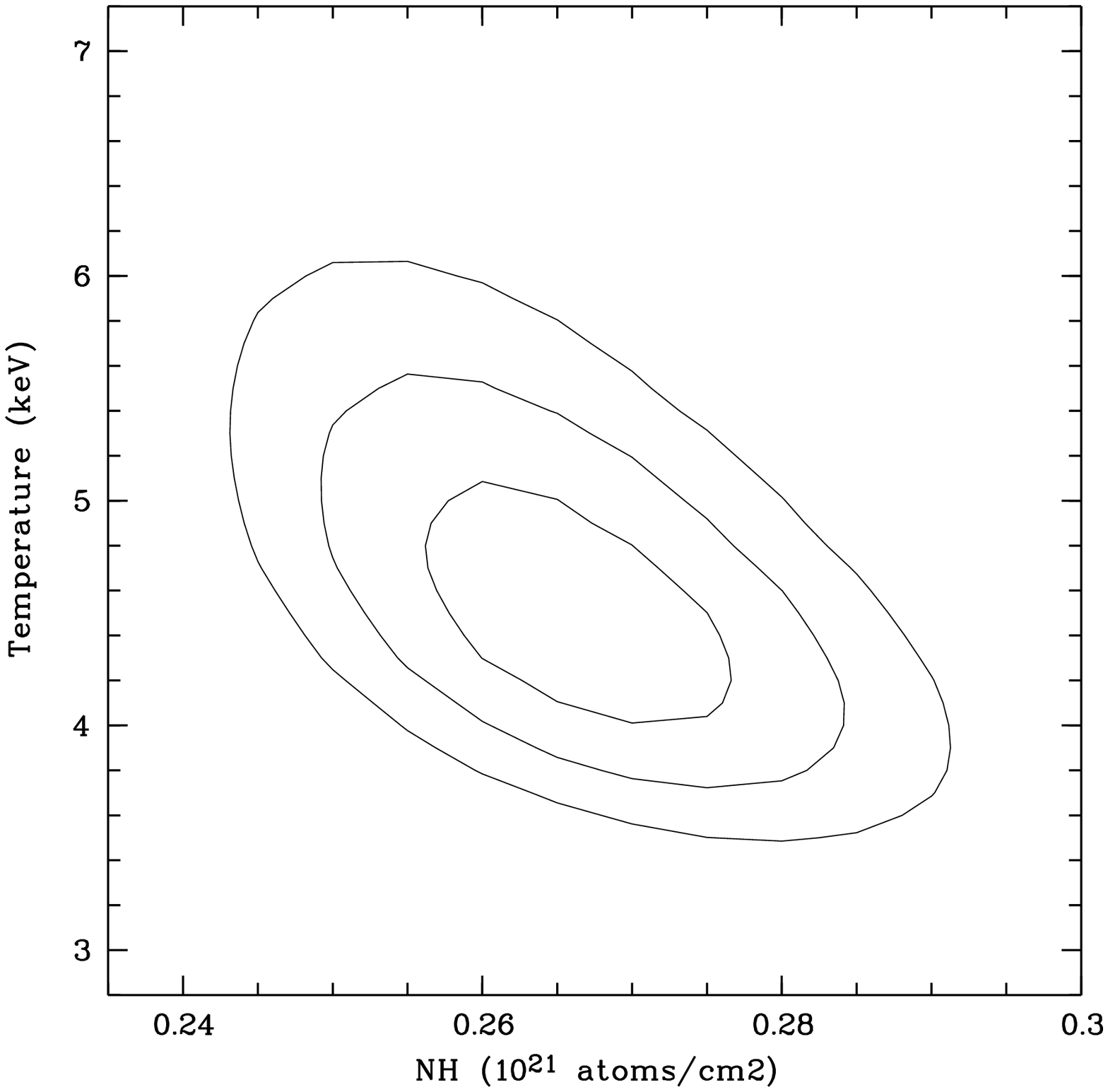,height=6cm,clip=on}}
        \caption{Contours of chi-squares values in the $N_{H}$ - T plane; the
metallicity is fixed at 0.2 solar. The contours are 1$\sigma$, 2$\sigma$, 
3$\sigma$.}
        \protect\label{contour1}
\end{figure}

\begin{description} 
\item[i)] A negative correlation 
between T and $N_{H}$. We have fixed the metallicity to Z=0.2 (in 
solar units, as assumed hereafter) and let T and $N_{H}$ vary as free 
parameters.  In Fig.~\ref{contour1} we show a contour plot of 
confidence levels 1, 2, and 3$\sigma$ of the $\chi ^2$ distribution 
around the central value N$_{\rm H}=2.7\ 10^{20}$cm$^{-2}$,
T=4.5~keV. 
\item[ii)] No correlation between T and metallicity, when we fix 
N$_{\rm H}= 3.58\ 10^{20}$~cm$^{-2}$, 
corresponding to the Galactic value (hereafter the ``canonical value'') 
deduced from HI data (Dickey \& Lockman 1990), and let T and Z vary as 
free parameters.  
\end{description} 

\subsection{Modelling of the X-ray gas}\label{vincent}
We have used the routines developed by Snowden et al. (1994) to determine the 
real exposure time, to subtract the various spurious background 
components, to correct for exposure, vignetting and variations of 
detector quantum efficiency and to merge the images.  A detailed 
discussion of the different PSPC background components can be found in 
Snowden et al.  (1994).  We obtain background subtracted images in 
four semi-independent energy bands for the image analysis between 
0.5 and 2.0~keV, chosen to have simultaneously a sufficient energy 
resolution and signal to noise ratio in each energy band, and to avoid strong
variations of the PSF with energy. The background we found is roughly 
$10^{-5}$ cts/sec/pix. We develop and improve a previous model 
which was used to fit X-ray images taken with the Einstein satellite (GDLL).  

We construct a ``synthetic'' image by 
assuming the X-ray emission due to thermal Bremsstrahlung, using for
the Gaunt factor the analytical approximation of Mewe et al.  (1986).  
We assume that the cluster has elliptical symmetry, and we take an axial ratio 
$\epsilon$ and the position angle of the major axis $\theta_{0}$ as 
free parameters; the ``radius'' $r$ we define is:
\begin{equation}
	 r^{2}=x^2 cos^2(\theta-\theta_0)
 +\frac{y^2}{\epsilon^2} sin^2(\theta-\theta_0)
 +\frac{z^2}{\epsilon^2}
	\label{er}
\end{equation}
assuming that the x and y axes are perpendicular to the line of sight, where
$\epsilon$ is linked to the ellipticity $e$ by $e=(1-\frac{1}{\epsilon}$).

For the electron density, we take a modified Hubble law ($\beta$-model), 
\begin{equation}
 n(r)=\frac{n_0}{{(1+(\frac{r}{r_c})^2)}^{\frac{3}{2}\beta}}
	\label{betamodel}
\end{equation} 
including a so-called ``analytical'' King profile obtained for $\beta 
=1$, and a modified Mellier-Mathez (Mellier \& Mathez, 1987, hereafter MM) law 
\begin{equation}
 n(r)=I_{0}\,(\frac{r}{a})^{-p}\,\exp[-(\frac{r}{a})^{\nu}]
	\label{gammamodel}
\end{equation} 
 
The free parameters are $\theta_{0}$ and $\epsilon$ for the geometry, $n_{o},
r_{c}$, $\beta$, and $I_{0}, $a$, \nu$ for these two laws respectively.  
The projection of the MM law leads to the Sersic law for the projected 
surface density $\mu$: 
\begin{equation}
\mu (r)=\mu _0\ e^{-(r/a)^\nu}
	\label{sersic}
\end{equation} 
when $p$ and $\nu$ are related by: $p\simeq 0.9976-0.5772\nu +0.0324\nu^2$
(Gerbal et al., submitted).

We multiply the volume-emissivity by the Galactic 
absorption factor (Morrison \& McCammon 1983), and then convolve the 
model with a Gaussian with FWHM equal to the energy uncertainty 
$\Delta\!E$ derived from the formula giving the spectral resolution: 
$\Delta\!E / E = 0.43\ (E/0.93\ {\rm keV})^{-0.5}$ (see e.g. Zimmerman et al. 
1994).
  
We then express the number of photons of energy E arriving on each 
pixel by integrating the number of counts along the line of sight, and 
sum over photon energies to obtain the 4 energy bands mentioned above.
In order to take into account the spatial resolution of the PSPC, we 
convolve our model with a two-dimensional Gaussian of FWHM = 25''.  
This approximation is valid only near the center of the detector, where 
the effect of the PSF on the true luminosity profile is more important.  
We could have taken the spatially dependent 1~keV on-axis PSPC point spread 
function (David et al., 1993), but because of the ``wobbling'' of ROSAT this 
method isn't more accurate.  

To compare with images created with Snowden's routines, we create images of 512 $\times$ 512 pixels (pixel size 14''.947 $\times$ 14''.947) 
and 256 $\times$ 256 pixels (pixel size 29''.894 $\times$ 29''.894) to 
increase the signal to noise ratio and to check that our fit is pixel 
size independent.

The best set of parameters is obtained by comparing the synthetic 
image to the observed one, pixel by pixel, using a maximum likelihood 
function to estimate this difference.  We assume that the number of 
counts in any pixel has a poissonian distribution.  The maximization 
of this function was performed with the MINUIT software from the CERN 
library.
\begin{figure}
	\centerline{\psfig{file=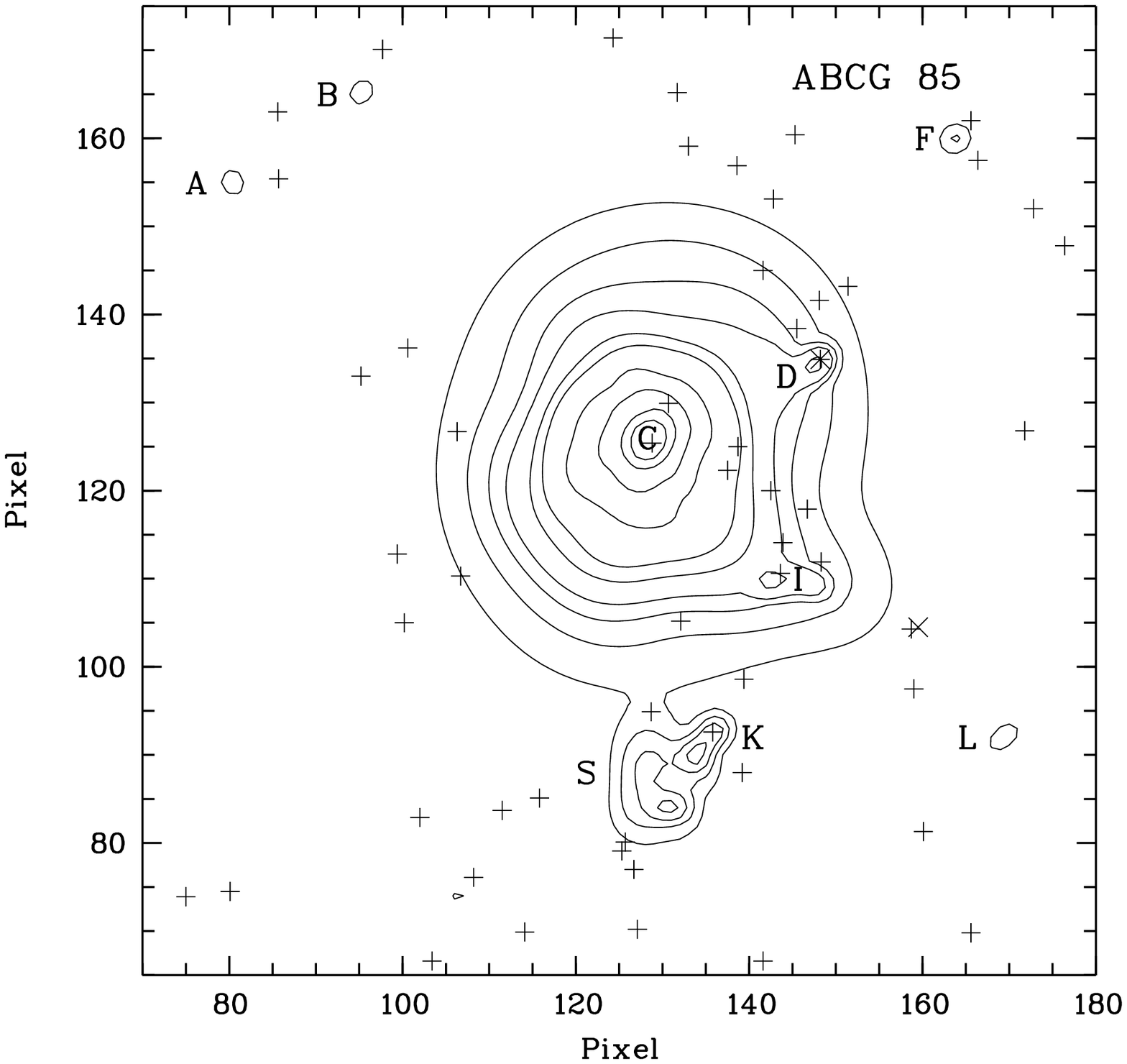,height=6cm,clip=on}}
	\centerline{\psfig{file=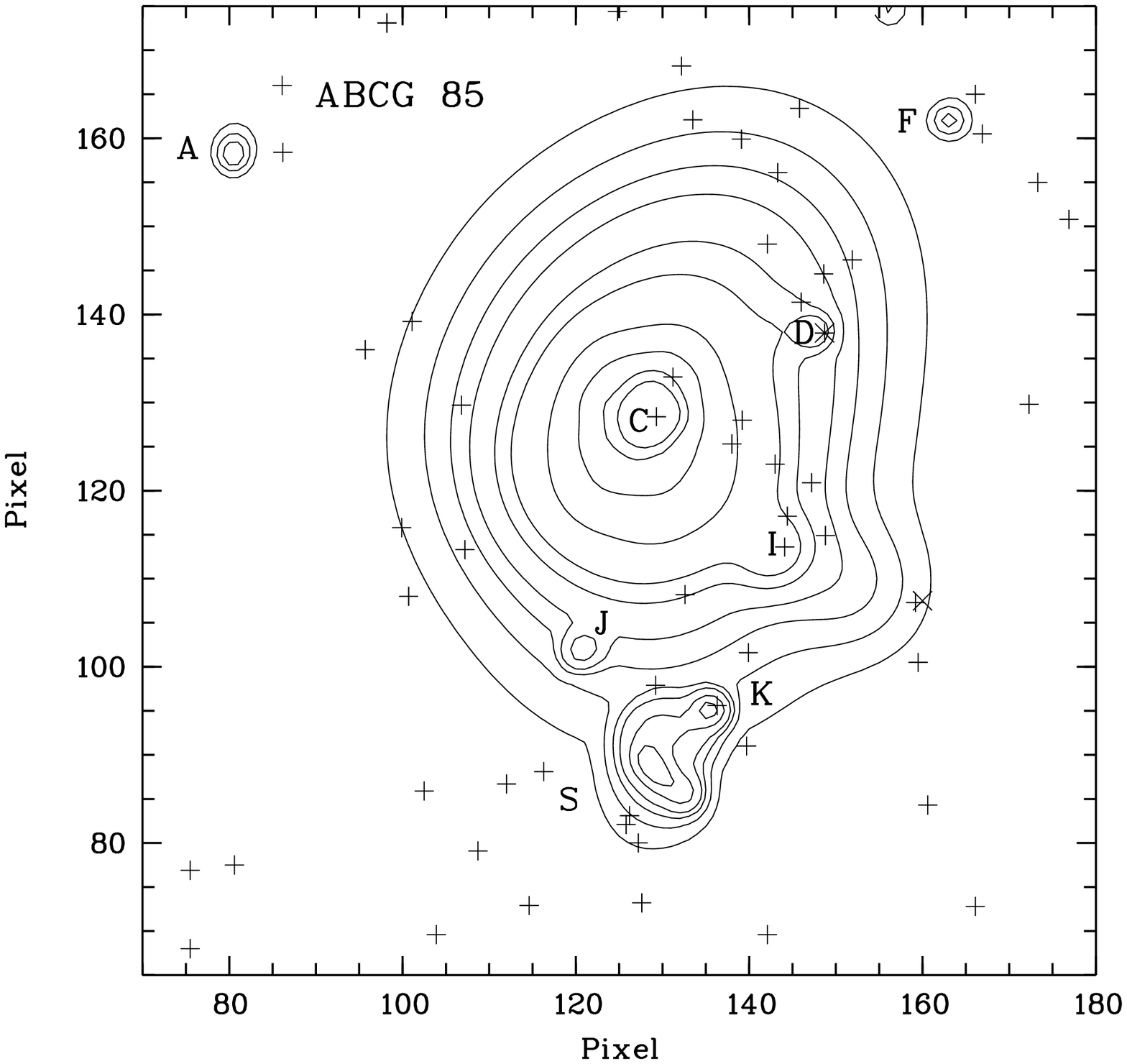,height=6cm,clip=on}}
	\caption{Reconstructed images. Top: German data, bottom: US data.
Values of the isophotes are: 1, 2, 3, 4, 6, 8, 16, 32, 64, 96, in raw counts.
Crosses indicate the positions of galaxies with redshifts belonging to the 
cluster (i.e. with velocities in the velocity range 13300-20000~km/s) and 
magnitudes brighter than R=17. Asterisks indicate emission-line galaxies.
North is top and East left, as in all following figures. The axes are scaled 
in ROSAT PSPC pixels of 15'' (23h$_{50}^{-1}$~kpc).}
	\protect\label{reconstruction}
\end{figure}
\section{Wavelet analysis results}\label{resultats}
\begin{figure}
	\centerline{\psfig{file=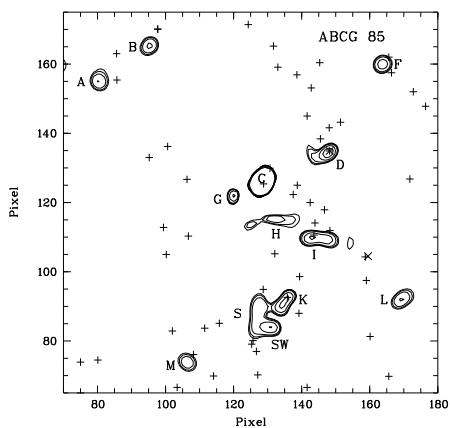,height=6cm,angle=-90}}
	\caption{German wavelet image at a scale of 2 pixels, or 30''
(see text for 
explanations). Values of the isophotes are 0.05, 0.1, 0.2, 0.5, 1 and 2 raw
counts. Same symbols as Fig.~\ref{reconstruction}.}
	\protect\label{coefficients}
\end{figure}

We now present the results of the wavelet analysis of both the German 
and US images, performed following the principles described in 
section~\ref{ondelettes}.  The reconstructed images where noise is 
suppressed are displayed in Fig.~\ref{reconstruction}.  Emission 
features with different spatial scales can now be disentangled by 
means of a classical wavelet analysis unfolding the data along a scale 
axis.  The German wavelet image at a scale of 2 pixels (plane 
2), i.e. 30'', is shown in Fig.~\ref{coefficients},  
emphasizing details of the image at the scale of the PSF.  The US wavelet 
image is very similar and will therefore not be presented here.    

Understanding the physical processes and dynamical situation in \a85 requires
a thorough analysis of these images.

The two reconstructed images present the same general aspect, a main 
structure and a substructure south of the main body (noted S), but some 
features differ from one image to another:
 \begin{itemize} 
 \item A feature (noted J) on the US image is not 
 present on the German image; another feature (noted B) is seen on the 
 German image but not on the American one.  It seems therefore likely 
 that these two substructures are artefacts in the raw data.  
 \item Substructure S is clearly of complex aspect and details
somewhat differ between the US and German images. 
A structure (noted K) located in the north-west part of blob S appears 
 elongated on the German image, while it has a globular aspect on the 
 US image.  
 \end{itemize} 

The central peak (noted C), visible in planes 1 and 2, and in the 
reconstructed images
is in fact unresolved, as noted in section~\ref{ondelettes}. The central pixel
contains $7\ 10^{-3}$ of the total number of counts within R$_{\rm L}$,
while its surface is $\sim 9\ 10^{-5}$ of the total surface within R$_{\rm L}$.

In a previous paper devoted to the Coma cluster of galaxies (Biviano et al.
1996), we have shown that wavelet images on the scales of 1 or 2 pixels 
allowed to find X-ray emitting galaxies after superposition of optical galaxy 
positions.  We have therefore superimposed on these images the 
positions of galaxies belonging to ABCG~85 (i.e. with heliocentric velocities 
in the interval 13300-20000~km/s).  We observe that the positions of several 
galaxies coincide exactly with significant features found at small scales in 
X-rays.  Other galaxies are  
close to X-ray substructures, but cannot be identified with certainty.  
We give the list of the identified galaxies and their characteristics 
in Table~\ref{galaxiesX}; the columns are the following: Col.~1: name, 
Col.~2: ROSAT coordinates, Col.~3 and 4: $\alpha$ and $\delta$ coordinates
(equinox 2000.0), Col.5~: velocity (cz) in km/s, Col.~6: R magnitude, Col.~7: 
quality of the galaxy 
identification, coded as: s=superposition of the galaxy position and 
X-ray substructure (within one ROSAT PSPC pixel, or 15''), c=close positions only (distance smaller than 3 ROSAT PSPC pixels).

In a companion paper devoted to the analysis of photometric and 
spectroscopic data for this cluster, we have applied the Serna-Gerbal 
method (Serna \& Gerbal 1996) which allows to separate dynamically 
structures and substructures in a field.  We have been able to show 
that in the field of ABCG~85 and in the velocity interval 
13300-20000~km/s, two structures can be found,
among which a ``Foreground'' structure poor in galaxies 
superimposed on the plane of the sky on a structure rich in galaxies: the \MS;
we also detect a background structure which will not be discussed here.
\begin{table*}[tbp]
\begin{minipage}{\linewidth}\centering
	\caption{List of X-ray emitting galaxies.}
	\begin{tabular}{rrrrccc}
	\hline
	Name & ROSAT       & $\alpha$~~~ & $\delta$~~~ & v & R & Quality \\
	     & coordinates & (J2000.0) & (J2000.0) &   &   & \\
	     & (pixels)    &          &          &   &   & \\
	\hline
C (cD)   &  128.8~125.4 & 0 41 50.07 & -9 18 10.41 & 16734 & 12.9 & s \\
D (A85-F) &  148.2~134.9 & 0 41 30.39 & -9 15 48.15 & 13429 & 13.9 & s \\
K        &  135.8~~92.6 & 0 41 43.05 & -9 26 22.32 & 16886 & 14.1 & s \\
H	 &  133.8~116.9 & 0 41 45.07 & -9 20 18.34 & 13831 & 17.0 & c \\
I        &  143.8~111.5 & 0 41 35.07 & -9 21 52.21 & 14234 & 15.4 & c \\
SW       &  129.5~~84.9 & 0 41 49.42 & -9 28 18.20 & 21647 & 17.8 & c \\
	\hline
		\hline
	\end{tabular}
	\protect\label{galaxiesX}
\end{minipage}
\end{table*}
The \MS ~has an average velocity of 16400~km/s and a dispersion of 
760~km/s (Durret et al. 1996).  In its geometrical and dynamical center it has 
a subset 
-- the \CSS -- of 8 bright galaxies (with R$<16.5$) that is dominated by 
the most luminous galaxy of the cluster (R= 12.9) and by the second 
most luminous galaxy of this subgroup (noted BCM$_{2}$) which has a 
magnitude R=14.1.  These two galaxies form a pair, while the 
other galaxies of this subset can be considered as their satellites.
>From a visual inspection of its profile, the central galaxy appears to be a 
cD, since it appears to have a point of inflexion in its surface 
brightness at $\mu _R\sim 23$, which looks like the onset of the cD 
halo (Colless, private communication). 

We notice that the X-ray image is far smaller than the galaxy map: in fact 
the \CSS ~covers all the X-ray map. The cD is situated just on the peak 
of X-ray emission, while the galaxy BCM$_{2}$ is located within 1 pixel of one
of the structures (K) in the north-west part of the S extension. 
Moreover a visual inspection indicates that the extension and 
orientation of BCM$_{2}$ correpond to the orientation of the feature K 
as observed on Fig.~\ref{coefficients}

The \CS ~has a mean velocity of 14100 km/s and a velocity dispersion of 
400~km/s.  It is located near the central part of the X-ray image. 
Assuming that the redshift of the Main and Foreground 
structures are due to the Hubble flow only, the distance between these 
two clusters would be 46~\h50 Mpc. It then appears difficult to 
consider the couple \MS--\CS ~as the elements of a double cluster.

The \CS ~contains X-ray emitters accounting for 
two remarkable features: the most luminous galaxy of this group 
coincides with feature D and is a Seyfert galaxy (A85-F in the CfA catalogue); 
the second luminous galaxy coincides with feature I. Both features are
seen on all wavelet planes and on the reconstructed image, and are 
responsible for the two west extensions that break the south-north symmetry of 
the main body of the X-ray image.
\begin{figure}
	\centerline{\psfig{file=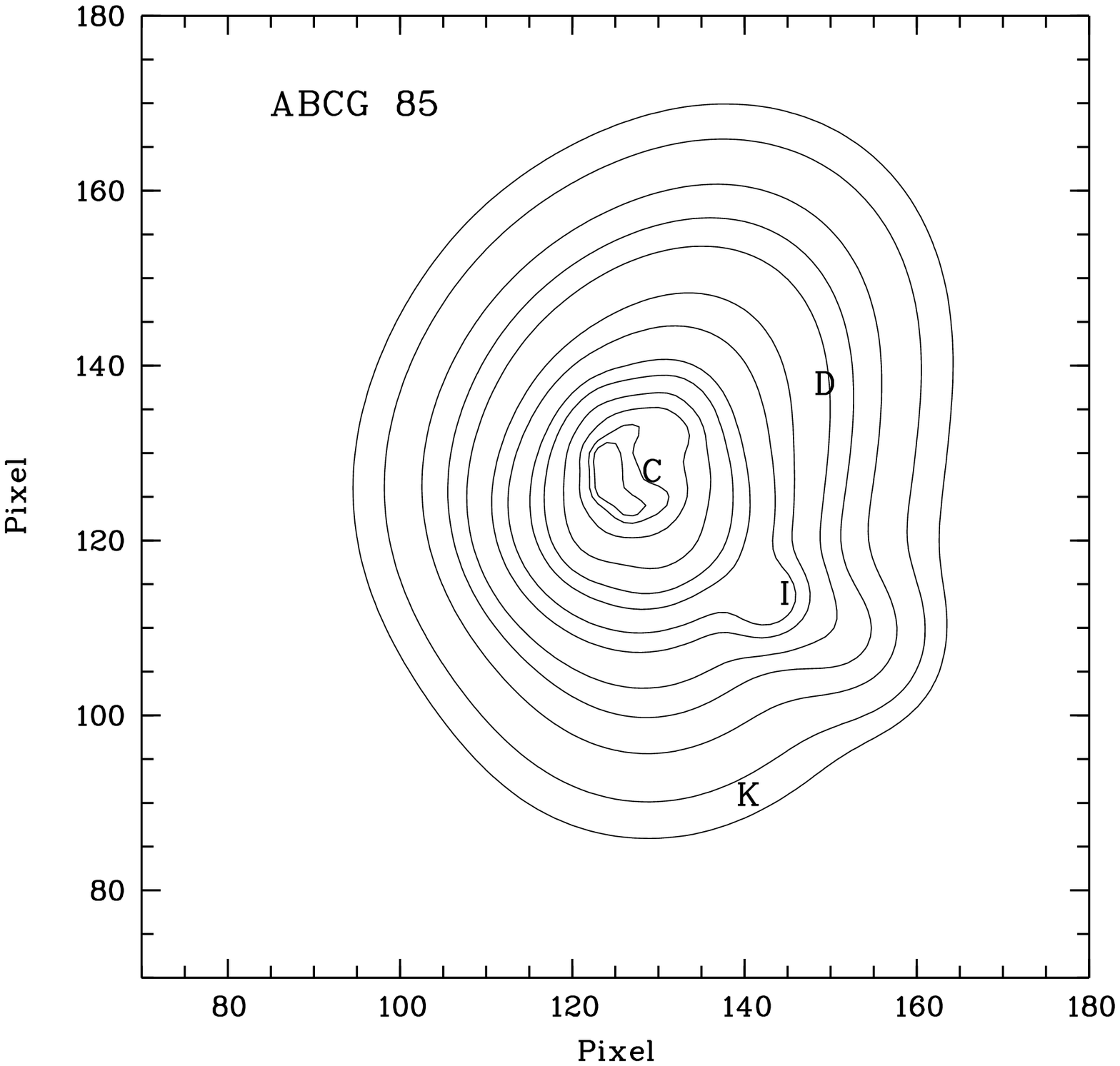,height=8cm,clip=on}}
	\caption{``Cleaned'' image, obtained after subtracting the features 
appearing on the wavelet images of planes 1 and/or 2 to the reconstructed 
image (see section~3). Values of the isophotes are 0.5, 1, 2, 
3, 4, 6, 8, 12, 16, 24, 36, 48, 54 raw counts. The letters are placed at the
positions of the various identified galaxies (see text).}
	\protect\label{nettoyee}
\end{figure}

We display in Fig.~\ref{nettoyee} the reconstructed US image from 
which point-like scale 1 and 2 objects have been suppressed.
This was done following the procedure described at the end of 
section~\ref{ondelettes}; this image, which only contains large scale
features, will hereafter be refered to as the ``Large scale image''.
This \large , showing the extended X-ray component (Fig.~\ref{nettoyee}), is 
much smoother and more regular than the reconstructed image. Its
examination leads to the following conclusions: 
\begin{itemize}
\item The central peak, which now appears as a hole (see 
section~\ref{ondelettes}) is not at the center of the extended image, 
but is displaced to the north-west and coincides with the position of the cD. 
This is not surprising, since cD galaxies are not necessarily in the very 
centre of clusters. It is possible but not proved that the central galaxy 
has an X-ray emitting nucleus which is responsible for this peaked 
feature.  
	
\item  The south extension S has disappeared, indicating that it is 
constituted by the superposition of small emission regions (such as K), and 
not by  a diffuse extended source such as hot X-ray emitting gas in a 
group of galaxies. 
\item  The D feature has vanished, showing the lack of an extended 
feature, as expected from the Seyfert nature of this galaxy. 
\item Contrary to the previous case the I feature has not 
disappeared, suggesting the presence of a diffuse component instead of a 
single X-ray emitting galaxy.
\item An extension is clearly visible towards the north-west.
\end{itemize}

\begin{figure}
	\centerline{\psfig{file=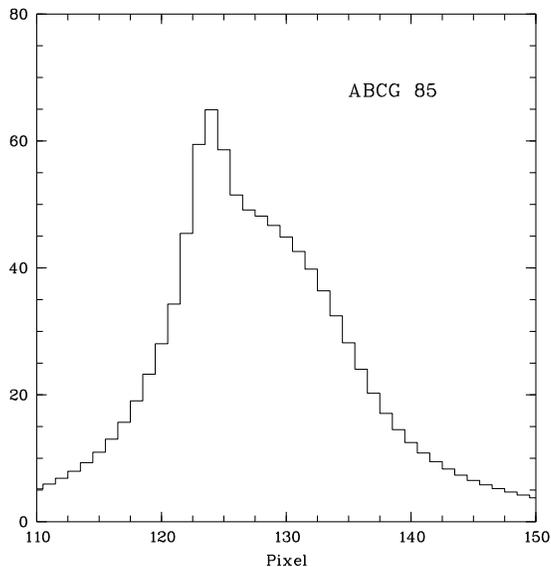,height=8cm,angle=-90}}
	\caption{Cut in the east-west direction of the ``cleaned'' image 
through the nucleus (defined as the pixel with the maximum number of counts). 
East is to the left and west to the right.}
	\protect\label{coupe}
\end{figure}
\begin{figure}
	\centerline{\psfig{file=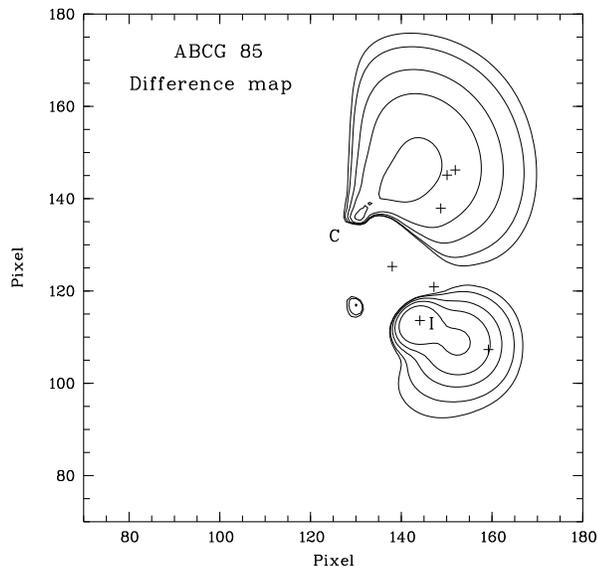,height=8cm,angle=-90}}
	\caption{Isophotes of the ``Difference'' image obtained as described 
in section~\ref{resultats}. Crosses indicate the positions of the galaxies in 
the ``Foreground Cluster'' mentioned in section~3. }
	\protect\label{difference}
\end{figure}

A cut through the center of the \large \ in the 
east-west direction (across pixel y=128) is shown in Fig.~\ref{coupe}.  
It shows a strong asymmetry, with a large excess of emission towards 
the west. This led us to construct a ``difference image'', using the following 
procedure: we cut the \large \ in the north-south direction; we 
kept the east half and added to it an image which was 
symmetrical to the east image, therefore 
obtaining an image with two perfectly symmetrical halves; we 
subtracted this symmetrical image to the \large, thus obtaining 
the ``difference image'' displayed in Fig.~\ref{difference}. The location 
of this difference image corresponds approximatively to that of the \CS.

In conclusion, the total image clearly corresponds to the superposition of
the X--ray emission by a group (\CS) superimposed on that of the central 
regions of a cluster (\MS) along the same line of sight, but 
separated by 46~\h50 Mpc.

\section{Spectral analysis results}\label{spectral}
Following the method described in section~\ref{spectre}, we have studied the 
temperature and its profile by taking into account the column 
density of hydrogen and the metallicity. 

\subsection{Global values}\label{GV}
\begin{figure}
	\centerline{\psfig{file=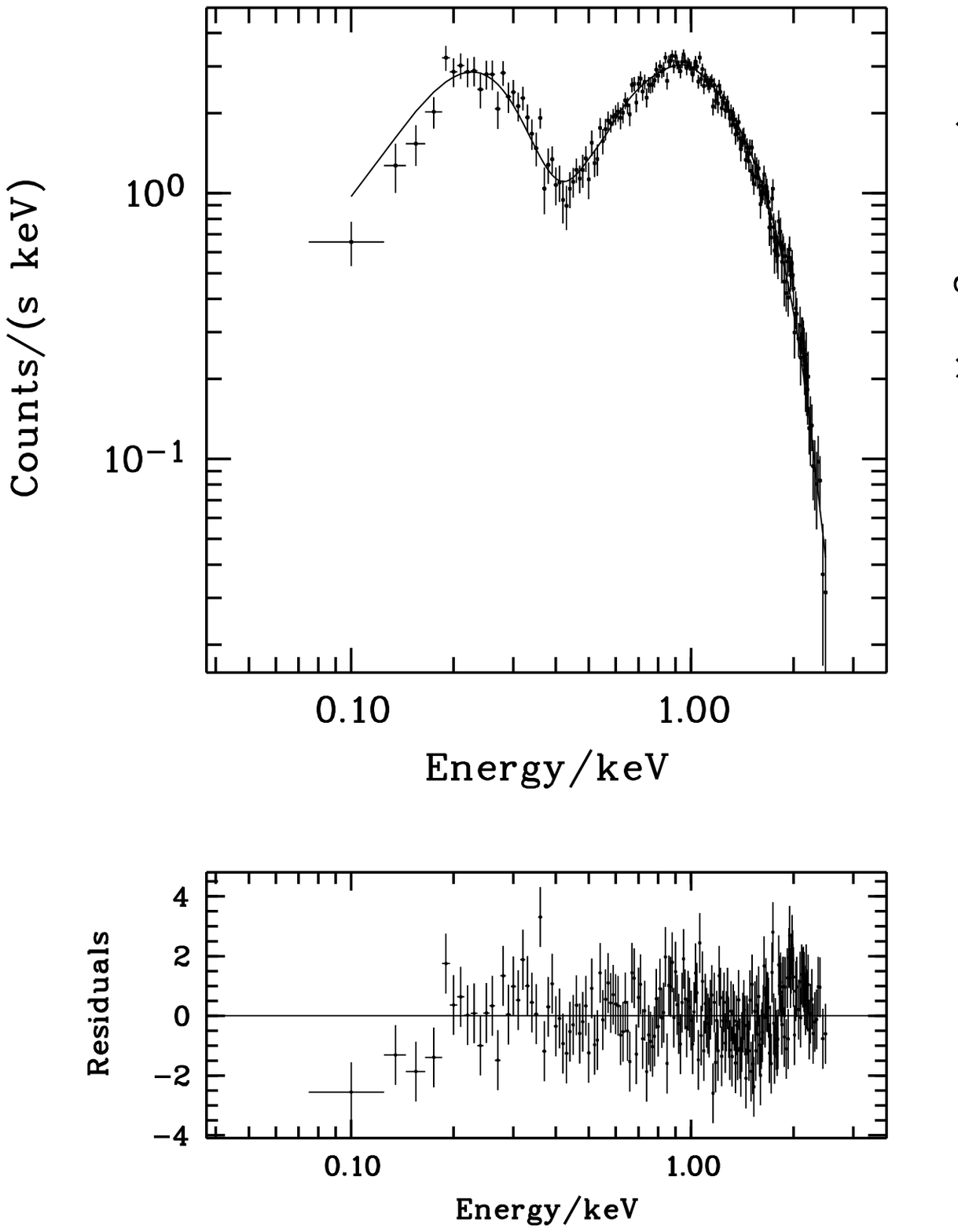,height=8cm,clip=on}}
	\caption{Background subtracted pulse height spectra of \a85 . The
metallicity is fixed to 0.3Z$_\odot$; the south blob is excluded. }
	\protect\label{spectrex}
\end{figure}
By using the Raymond \& Smith model applied globally to all the 
cluster, leaving free the column density as well as the temperature 
and metallicity, we find a temperature of T= 4.5$\pm 1.0$~keV, a 
column density N$_{\rm H} = (2.7 \pm 0.2)10^{20}$~cm$^{-2}$ and a 
metallicity Z=0.22$\pm 0.17$Z$_\odot$.  The temperature becomes T= 
4.6$\pm$0.7~keV, with a column density N$_{\rm H}= (2.6\pm 0.2)\ 
10^{20}$~cm$^{-2}$, when the metallicity is fixed to Z=0.3Z$_\odot$
(see Fig.~\ref{spectrex}). If we fix 
the hydrogen column density to the Galactic value of N$_{\rm H}= 3.58\ 
10^{20}$~cm$^{-2}$, we find a temperature of $3.4 \pm 0.3$~keV.  
However, we can notice that in the first two cases the column density 
of hydrogen is weaker than the canonical value; although this is the best 
mathematical solution, it appears physically doubtful, though
not totally excluded by the 1$^\circ$ mesh of the neutral hydrogen mapping.
In any case, all these results are consistent with a temperature 
T$\sim 4\pm 1$~keV.

A strong correlation between the temperature of the X-ray gas and the 
dispersion of galaxy velocities, expected since these two components 
lie in the same gravitational potential, has been noted for a long 
time (Mitchell et al.  1977). However, the actual value of the linear 
regression is still controversial (see for example the discussion 
by Edge \& Stewart 1991, based on EXOSAT data).  A small value of the 
velocity dispersion is expected from the temperature that we obtain.  
The analysis we have performed on our catalogue of velocities 
indeed leads to a main cluster with a velocity dispersion of 
760~km/s.  Note that the temperature that we find is notably lower 
than that given by David et al.  (1993) in their catalogue (6.2~keV), 
but derived using the MPC detector of Einstein, and also lower than 
that obtained by GDLL in their analysis of Einstein archive data 
($9\pm 4.5$~keV). To our knowledge, no temperature has been derived for this
cluster from ASCA data.

To a temperature of 4~keV and a Galactic column density corresponds an X-ray 
luminosity within R$_{\rm L}$ in the [0.1-2.4~keV] 
band of $9.3\pm 0.2\ 10^{44}$~erg~s$^{-1}$ for the main cluster (blob S being 
excluded). We would have liked to estimate the X-ray luminosity of the
``difference'' image, but this is not possible because we do not have
enough photons to determine the temperature. We only measured the total number 
of counts, which is 71000 for the total image, 64600 for the double of the
east image and 6400 for the difference image (background excluded).
Notice that the contribution of the difference image to the total X-ray counts 
is only $\sim 10\%$, while the luminosity  of the difference image in the R
band is $\sim 20\%$ of that of the cluster (see section~\ref{massdi}). This is
consistent with a lower temperature for the foreground group and with the
velocity dispersion of $\sim 400$~km~s$^{-1}$ found for this foreground group.

\subsection{Profiles}
We now analyse the temperature, column density and metallicity 
profiles.  To construct these profiles, we have defined annuli of 
variable widths; these widths were chosen to have a sufficient number 
of photons in each annulus for a significant analysis ($\geq$5000 photons 
in the central zone and outermost region and $\geq$10000 photons in the other annuli, after
background subtraction).

\begin{figure}
	\centerline{\psfig{file=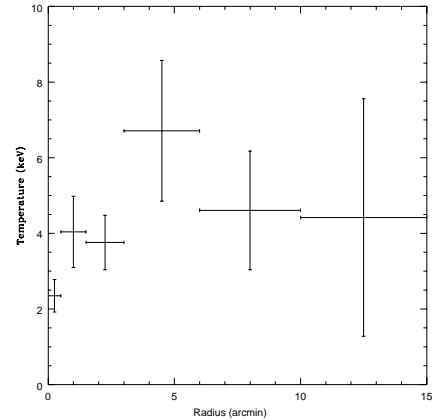,height=6cm}}
	\caption{Temperature profile obtained when the hydrogen column density
is a free parameter and the metallicity is fixed to Z=0.3Z$_\odot$.}
	\protect\label{trnhlib}
\end{figure}
\begin{figure}
	\centerline{\psfig{file=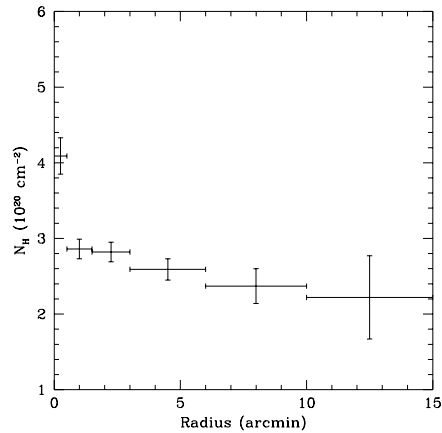,height=6cm}}
	\caption{Profile of the hydrogen column density with
the metallicity fixed to Z=0.3Z$_\odot$.}
	\protect\label{nhr}
\end{figure}
\begin{figure}
	\centerline{\psfig{file=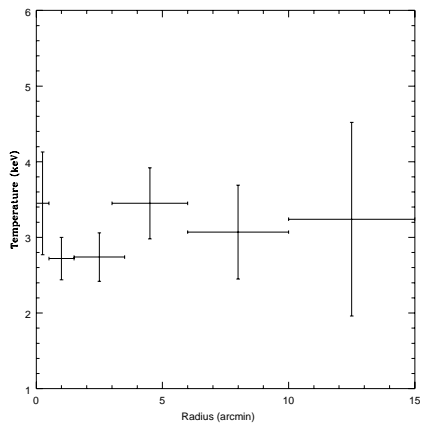,height=6cm}}
	\caption{Temperature profile obtained when the hydrogen column density
is fixed to its canonical value (the Galactic value) and the metallicity is 
fixed to Z=0.3Z$_\odot$.}
	\protect\label{trnhfix}
\end{figure}

In the profile shown in Fig.~\ref{trnhlib} both the temperature and 
hydrogen column density were free parameters, while the metallicity 
remained fixed to Z=0.3Z$_\odot$.  We observe in this case a temperature which 
is smaller in the central regions, while annuli located further out 
show a higher temperature. This temperature profile is consistent with
that of Kneer et al. (1996). Fig.~\ref{nhr} shows that the column 
density has an inverse behaviour: a much higher value in the very center, and 
smaller values in the outer regions, even unacceptably smaller than the 
lower limit N$_{\rm H}= 3.58\ 10^{20}$~cm$^{-2}$.  Such a behaviour 
could be due to the negative correlation between the X-ray gas 
temperature and N$_{\rm H}$ (see Fig.~\ref{contour1}).  We have 
therefore fixed hereafter the column density to this canonical value.  
The temperature profile, displayed in Fig.~\ref{trnhfix}, is then 
compatible with an isothermal profile. The data and fitting methods that we 
use forbid an objective choice between both temperature profiles. Note that
excess absorption in the very centre has been reported by Prestwich \& Daines 
(1994) from ASCA data; it could be interpreted as due to neutral hydrogen in 
the cD galaxy and/or to one of the cold phases of the intracluster gas.

We have searched for a variation of metallicity with radius, by 
letting the metallicity and temperature vary in concentric regions, 
leaving N$_{\rm H}$ fixed to its canonical value. 
In the very central part 
(radius of 30~arcsec), where the cD is located, we find a high metallicity 
Z$=0.83\pm 0.31$Z$_\odot$ and a temperature T=$2.9\pm 0.6$~keV. In all the other concentric annuli (or their sum), we only obtain an upper 
limit of $\sim$0.1Z$_{\odot}$ 
for the metallicity. The metallicity is therefore significantly larger in the
center than the usual value of Z$\sim 0.3$Z$_{\odot}$. Moreover, other fits
with various values of N$_{\rm H}$ (even if N$_{\rm H}$ is left free to vary) 
always lead to higher values of Z in the center.
Further out we only have upper limits for the 
metallicity; this is consistent with the global value Z$\sim 0.2$Z$_\odot$ 
derived above (section~\ref{GV}). The large difference in metallicity that we 
find between the center and other regions is comparable to the steep 
metallicity gradients respectively found for the Virgo and Centaurus clusters 
by Koyama et al. (1991) using GINGA and Fukazawa et al.  (1994) using ASCA. 

\section{Modelling results}\label{ajustements}
The description of the morphology of the cluster presented in 
section~\ref{resultats} shows the difficulty of applying the 
fitting method described above (section 2.4).  In particular, the fact that the
south blob is the sum of independent objects forbids us to fit this 
region assuming a diffuse distribution of X-ray gas. We therefore excluded 
from our analysis the angular sector between $\theta = 135^{o}$ and 
$\theta = 225^{o}$ (anticlockwise from north) inside which this secondary 
component lies. Taking into 
account the fact that the contribution of what we have called the \CS ~
represents $\sim 10\%$ of the total number of counts in the image, we have 
neglected it in our analysis.  The central peak is displaced relatively to the 
center of the large scale configuration; however the distance between 
this peak and the center is small (two or three pixels); we have therefore 
modelled the image by situating the center of our synthetic image on this peak.
Our analysis was applied to images of 
512 $\times$ 512 and 256 $\times$ 256 pixels. For the sake of simplicity,
the temperature was assumed to be isothermal with a value of 4~keV. 
An attempt to fit the contribution of the central peak has not given 
convincing results.

\subsection{Density profiles}\label{densprof}
\begin{figure}
	\centerline{\psfig{figure=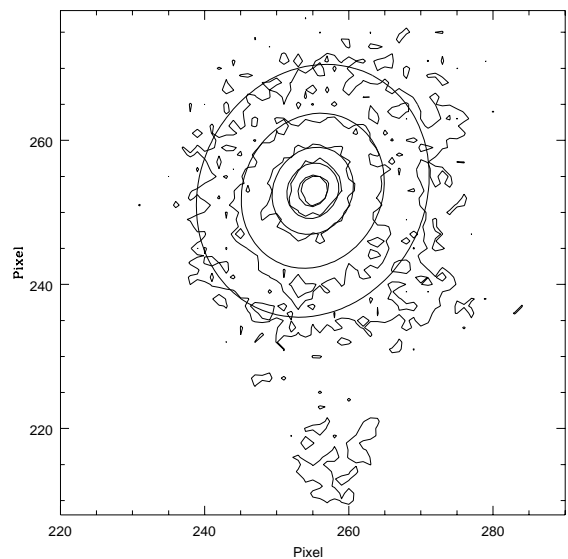,height=8cm}}
\caption[ ]{Isopleths of the central region of ABCG~85 with those 
of the synthetic image obtained
with the best modified Hubble model fit superimposed. The $90^{o}$ angular 
sector between $\theta = 135^{o}$ and $225^{o}$ inside which the secondary 
component lies, was excluded from the fit. The fit was performed until 15'.
The pixel size is 0.25 arcmin.}
\protect\label{fit}
\end{figure}
\begin{table*}[tbp]
\caption[ ]{Results for ABCG~85. Modified Hubble density profile.}
\begin{tabular}{cccccccccc}
\\
\hline
Model &$n_{o}$&$\beta$& $r_{c}$ & Gas mass & Gas mass & Gas mass & 
Dyn. mass & Dyn. mass & Dyn. mass \\
  &($10^{-3}{\rm cm}^{-3}$)& & ({\rm kpc}) & (1 Mpc) & R$_{\rm L}$ & (3 Mpc) 
& (1 Mpc) & R$_{\rm L}$ & (3 Mpc) \\
  & & &  & ($10^{13} M\odot$) & ($10^{13} M\odot$) & ($10^{13} M\odot$) &
($10^{14} M\odot$) & ($10^{14} M\odot$) & ($10^{14} M\odot$) \\
\hline
1 &6.5&0.497& 60.0 & 1.7 & 3.1 & 9.4 & 1.5 & 2.2 & 4.6 \\
3$\sigma$ errors &0.4& 0.006 & 3.7 & 0.4 & 0.7 & 2.0 & 0.5 & 0.7 & 1.5 \\
2 &10.8&0.438& 32.1 & 2.1 & 3.9 & 13.4 & 1.5 & 2.1 & 4.4 \\
3$\sigma$ errors &1.0& 0.005 & 3.1 & 0.6 & 1.1 & 3.9 & 0.5 & 0.7 & 1.4 \\
3 &   &1.000& 470  &     &      &     &    \\
\hline

\end{tabular}

\noindent
1 : pixel size = 0.5 arcmin \\
2 : pixel size = 0.25 arcmin \\
3 : King profile \\
\end{table*}
\begin{table*}[tbp]
\caption[ ]{Results for ABCG~85. Modified MM density profile.}
\begin{tabular}{ccccccccccc}
\hline
Model & $I_{o}$ & $\gamma$ & $a$ & Gas mass & Gas mass & Gas mass & Dyn. mass
& Dyn. mass & Dyn. mass \\
  & ($10^{-3}{\rm cm}^{-3}$) &  & ({\rm kpc}) & (1 Mpc) & R$_{\rm L}$ &
 (3 Mpc) & (1 Mpc) & R$_{\rm L}$ & (3 Mpc) \\
  &  &  &  & ($10^{13} M\odot$) & ($10^{13} M\odot$)  & ($10^{13} M\odot$)
& ($10^{14} M\odot$) & ($10^{14} M\odot$) & ($10^{14} M\odot$) \\
\hline
4 & 1.5 & 0.62 & 430 & 2.8 & 4.8 & 11.5 & 1.7 & 2.9 & 8.3 \\
3$\sigma$ errors & 0.06 & 0.02 & 14 & 0.5 & 0.8 & 2.0 & 0.5 & 0.9 & 2.7 \\
5 & 3.5 & 0.327 & 190 & 2.8 & 5.1 & 16.2 & 1.5 & 2.3 & 5.3 \\
3$\sigma$ errors & 0.8 & 0.025 & 37 & 2.0 & 3.6 & 11.3 & 0.5 & 0.7 & 1.7 \\
\hline

\end{tabular}

\noindent
4 : pixel size = 0.5 arcmin; MM profile\\
5 : pixel size = 0.25 arcmin; MM profile \\
\end{table*}

\begin{figure}
\centerline{\psfig{figure=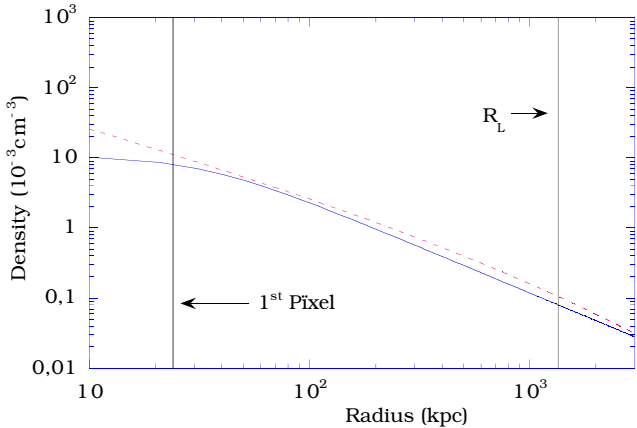,height=6cm,clip=on}}
\caption[ ]{X-ray gas density profile; full line: modified Hubble law, 
dashed line: MM law (values are taken from the fits with a pixel size of 
0.25 arcmin). The size of the first pixel is indicated, as well as the 
maximum radius of the image (R$_{\rm L}$). }
\protect\label{profildens}
\end{figure}

The fits are performed up to the limiting radius R$_{\rm L}$ (15'), or 
1.4~\h50~Mpc, corresponding to the radius where the zero-level is reached.
However, we find no significant change if the fit is extended up to a 
radius of 18'. The best fit gives an ellipticity $e=0.14\pm 0.02$, with the 
major axis located along PA=152$^\circ$. The parameters of the fits as well 
as other physical quantities derived from our fitting procedures are 
summarized in Tables~2 and 3.  The isophotes of the PSPC image and of a 
typical synthetic image are displayed in Fig.~\ref{fit}. $\beta$-model and MM 
profiles are shown in Fig.~\ref{profildens}: they are identical, except in 
the central pixel.

Our analyses provide simultaneously small core radii and weak slopes. 
A similar result is obtained when we construct a gas profile from 
concentric elliptical annuli, leading to $\beta =0.46\pm 0.02$ and 
$r_c=55\pm 8$~kpc. Note that these two parameters are 
correlated; this explains why the core radius obtained with a King profile 
($\beta$=1) is large (however in this case the quality of the fit is poor). 
These results are consistent with those obtained from Einstein data 
using a comparable procedure (GDLL, Durret et al.  1994). They also agree
with those found by Matilsky et al. (1985), Schindler et al. (1996) and 
Schindler \& Wambsganss (1996) for other clusters, but not with the larger 
values derived by Jones \& Forman (1984). 
A bad background subtraction could lead to low values of $\beta$; however,
our background subtraction was performed using the software developed by
Snowden et al. (1994) and we have checked that it was correctly done. 
Besides, analyses in the central region of clusters acting as gravitational 
lenses, have shown that the dark matter distribution would present extremely 
small core radii.  This set of results --~although \a85 is not known as a 
lense cluster~-- strenghtens our confidence in our results.

\subsection{Mass distribution}\label{massdi}
We integrate the density profiles determined above to obtain the 
mass of the X-ray gas. The dynamical mass was derived under the assumption of 
hydrostatical equilibrium, using the following formula:
\begin{equation}
M_{dyn}(r) = \frac{3.7\,10^{10}}{\epsilon ^2}\,r^{2}\,T(r)\,\frac{d}{dr}[ln(n(r)T(r))]
	\label{masse}
\end{equation} 
\noindent 
where $r$ is in kpc, $T$ in keV and $M_{dyn}(r)$ in solar masses.

\begin{figure}
\centerline{\psfig{figure=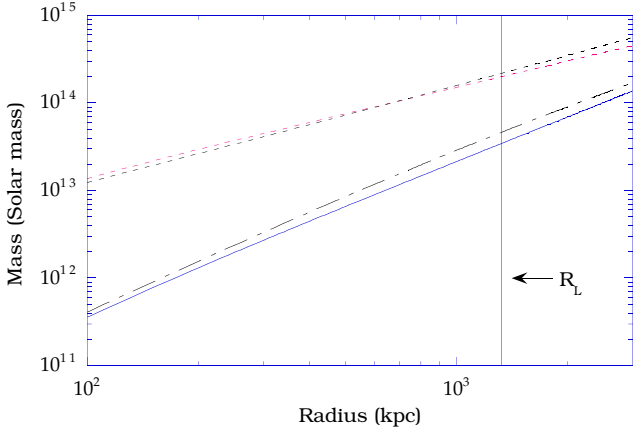,height=6cm,clip=on}}
\caption[ ]{X-ray gas mass (full line: modified Hubble law, dot-dashed line:
MM law) and dynamical mass (dashed line: modified Hubble law, bold
dashed line: MM law (values are taken from the fits with a pixel size of 
0.25 arcmin). The maximum radius of the image is indicated (R$_{\rm L}$). }
\protect\label{masses}
\end{figure}

The masses were computed at the distances $1h_{50}^{-1}$ Mpc, R$_{\rm L}$ and 
also at $3h_{50}^{-1}$~Mpc (even if 3~Mpc is out of the image) to compare 
with previous studies (GDLL, Jones \& Forman 1984).  Results are given 
in Tables~2 and 3 for a Hubble constant H$_\circ$=50~km~s$^{-1}$~Mpc$^{-1}$. 
The dependences of these masses with H$_\circ$ are:
M$_{X-ray gas}\propto h_{50}^{-5/2}$ and
M$_{dyn}\propto h_{50}^{-1}$. The gas and dynamical masses found here are 
smaller than
previously found (GDLL) by factors of 2 and 3 respectively; this can be 
accounted for by the fact that the dynamical mass depends on the X-ray gas 
temperature, which is now taken to be lower. Henriksen \& White (1996) give 
larger gas masses but inside a radius of 4.4~Mpc, a distance to which an 
extrapolation is hazardous.  Integrated mass profiles are shown in 
Fig.~\ref{masses}.  

\begin{figure}
\centerline{\psfig{figure=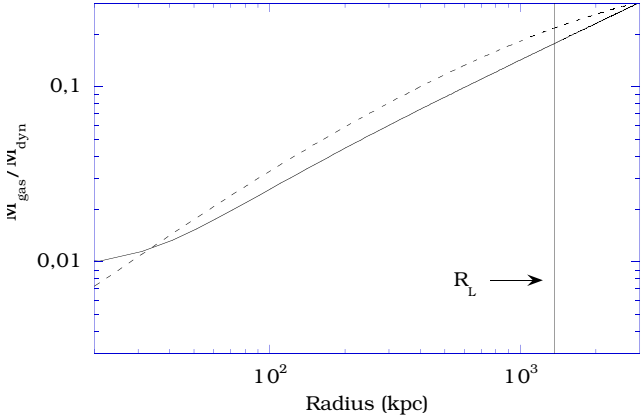,height=6cm,clip=on}}
\caption[ ]{Ratio of the X-ray gas to dynamical masses; full line:
modified Hubble law, dashed line: MM law (values 
are taken from the fits with a pixel size = 0.25 arcmin). }
\protect\label{ratio}
\end{figure}
The ratio of the X-ray to dynamical masses is shown in Fig.~\ref{ratio}.  
The very steep slope of this curve indicates that the binding matter is much 
more centrally condensed than the gas, as found in previous studies based on 
Einstein data (see e.g.  Durret et al.  1994 and references therein).  The 
behaviour is the same for both density profiles. We find that the 
X-ray emitting matter (up to $R_{L}$) is $(18\pm 4)\%$ of the 
dynamical mass.  The fact that $\beta$ is found to be smaller than 1 
leads to a divergent gas mass for large radii, and to a ratio 
M$_X$/M$_{dyn}>1$ for a sufficiently large radius.  The existence of a 
strong decrease of the density at large radius is therefore expected to be 
observed by future X-ray satellites with higher sensitivities.

Using our optical data, we have calculated the luminosity in 
the V band in the same region as the X-ray image for galaxies with
redshifts in the cluster. We find a total luminosity of
$L_{\rm V}\sim 8.4\ 10^{11}h^{-2}L_\odot$, leading to a mass-to--light ratio
$ M/L_{\rm V}\sim 300$ in solar units. Notice that we have taken into account 
in our calculation the light coming from the \CS; the luminosity originating 
in this group can be estimated as roughly $20\%$ of the total luminosity
(estimated in the V band). Assuming a stellar mass-to-light ratio 
$M_*/L_{*}= 8$ (in solar units), the corresponding stellar mass is 
$\approx 6.7\ 10^{12}h_{50}^{-2}\ M_{\odot}$, which is not totally negligible
compared to the gas mass, since $M_*$ is about 20$\%$ of M$_{\rm X}$. 
Therefore, roughly 83\% of the matter in this cluster is unseen matter.

\section{Discussion and Conclusions}\label{chieze}
\subsection{Sunyaev-Zel'dovich effect}\label{szeffect}
The density distribution of the X-ray gas derived from our model can be
used in the computation of the Sunyaev-Zel'dovich effect in the direction
of this cluster. We derive the relative temperature decrease of the microwave
background temperature $T_\gamma(r)$ in the direction of \a85 by applying the 
usual formula (see e.g. Liang 1995):
\begin{equation}
\frac{\Delta T_{\gamma}(r)}{T_{\gamma}(r)} = 
\frac{-2\sqrt{\pi}\sigma _{T}kT}{m_ec^2}\ n_{2D}(r)
	\label{EsseZed}
\end{equation} 
where $T$ is the X-ray gas temperature, $\sigma _T$ the Thomson scattering 
cross-section, $m_e$ the electron mass and $ n_{2D}(r)$ the gas density 
projected on the plane of the sky.
\begin{figure}
\centerline{\psfig{figure=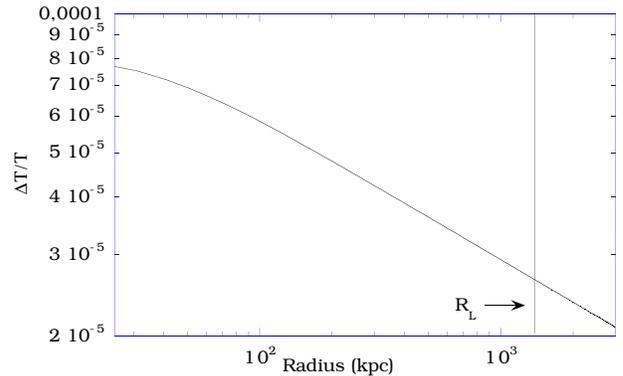,height=5cm,clip=on}}
\caption[ ]{Relative temperature decrease in the microwave background radiation
due to the Sunyaev-Zel'dovich effect in the direction of \a85.}
\protect\label{sz}
\end{figure}
The corresponding curve showing the relative decrease of the microwave
background radiation temperature in the direction of \a85 is displayed in
Fig.~\ref{sz}. The central value is small ($\simeq7~10^{-5}$) and it does not seem possible 
to observe it with present techniques, but \a85 is a cluster typical of
those that the next generation of submillimeter waveband satellite 
observatories such as Cobras/Samba 
will observe (Bersanelli et al. 1996).
\subsection{Accretion flow and metallicity}\label{ct}
\a85 is a cluster reputed to have a strong central cooling flow
(Prestwich et al. 1995, and references therein). The main argument
in favour of a cooling flow in \a85 is the fact that the cooling time,
$t_{\rm cool} \approx (d \ln E/dt)^{-1} \propto n^{-1} T^{1/2}$ is smaller than
the cluster age.
We may suppose that the gas has lost a significant
amount of energy in this region and that mass deposit is taking place.

Descriptions of the distribution and temperature of the X-ray gas based on
the formalism of hydrodynamics (or the wind formalism) have been undertaken 
long ago, at the time of the first X-ray observations. The structure of 
equations is formally given by:
\begin{equation}
\mathcal{D}[\mathrm{A}]=-\sigma + \mu
	\label{transport}
\end{equation} 
\noindent
$\mathcal{D}$ describing a transport operator applied to [A], which can
represent the density, momentum or energy.  $\sigma$ represents the 
``sinks'', which in the energy equation will account for radiation 
losses, and in the density equation will account for losses by matter 
deposit.  $\mu$ represents the injection of matter, momentum and heat 
originating for example from galaxy winds (generated for instance by 
super-nova explosions).  These equations very often present two types 
of steady-state solutions: Yahil \& Ostriker (1973) gave ``outflow'' solutions,
and Cowie \& Binney (1977) proposed an accretion flow model, where the flow is
regulated by radiation, in order to account for the fact that the cooling 
time is of the order of the age of the cluster, and for a possible central 
over-density.  They predicted the formation of clouds and of stars, i.e.  a 
large accretion of gas onto the central galaxy.

The resolution of these equations requires 
the knowledge of the gravitational potential.  It is possible to calculate 
it to a good approximation from the resolution of the hydrostatic equation, 
then in a second step to use this potential to obtain accretion
models;  the deposit is supposed to take place only in the 
region where the cooling time is smaller than the cluster age
(Fabian et al. 1984); it is this 
last type of analysis which is at the basis of the cooling flow paradigm. 
Using the standard steady sub-sonic flow model approximation (Fabian \& Nulsen 
1977) we estimated the rate of mass deposit $\dot{M}$ in the central region of 
\a85. For all our models, we estimated the rate of mass deposit for the two
extreme cluster ages $1\ 10^{10}$~years and $2\ 10^{10}$~years. The results
are displayed in Table~4. The mass deposit rates range from $19\pm 12$ 
to $68\pm 38$~M$_\odot$/yr. 
These values are lower than those derived by Stewart et al. (1984),
who found for this cluster $\dot{M} \approx$ 100 M$_{\odot}$/yr from
Einstein IPC data, or Edge et al. (1992) who give $\dot{M} \approx$ 236 
M$_{\odot}$/yr from Einstein MPC data; however, our values agree 
with the average value given by Godon et al. (1994). Differences mainly 
come from a much larger core radius obtained by the first two sets of authors 
(the effects of differing values of $\beta$ and of the temperature are less 
important).

\begin{table}[tbp]
\caption[ ]{Cooling radius and mass deposit computed for the 4 models 
described in the text, and for cluster ages $1\ 10^{10}$ and 
$2\ 10^{10}$~years.}
\begin{tabular}{cccc}
\\
\hline
Model  & t$_{\rm cool} (10^{10}$ yr) & r$_{\rm cool} {\rm (kpc)}$ & 
$\dot{M} (M_\odot$/yr) \\ 
\hline
 1    &    1         &      $   25 \pm ~4  $ &  $  25 \pm 14$ \cr
      &    2         &      $   84 \pm 14 $ &  $  54 \pm 30$ \cr
 2    &    1         &      $  41 \pm ~7   $ &  $  29 \pm 17$ \cr
      &    2         &      $  82 \pm 14  $ &  $  42 \pm 25$ \cr
 3    &    1         &      $  38 \pm ~6   $&   $ 25  \pm 15$ \cr
      &    2         &      $  88 \pm 15  $ &  $  68 \pm 38$ \cr
 4    &    1         &      $  36 \pm ~6   $ &  $  19 \pm 12$ \cr
      &    2         &      $  79 \pm 13  $ &  $  49 \pm 27$ \cr
\hline
\end{tabular}
\end{table}

This calculation neglects the injection of gas, momentum and energy 
in the intra--cluster medium. i.e. assumes that the $\mu$ 
term in equation~\ref{transport} vanishes. 

 Since the non steady-state models presented by Hirayama et al.  
 (1978), numerous studies have been done to describe the formation of 
 clusters from an initial perturbation.  One of the great difficulties is 
 the crucial role played by the cooling function of the gas 
 (Blanchard et al. 1992).  As a consequence, the existence of a 
 multiphase medium appears unescapable.  Transport of gas from  hot phases to 
 colder phases occur, and the consequence of such a cascade is the 
 formation of stars.  It is a precise description of this process (in 
 fact the same as in the insterstellar medium) which accounts for the $\sigma$ 
term in 
 Equation~\ref{transport} (Chi\`eze et al., in preparation).  The 
 next step is, re-injection of metal-rich gas and energy into the 
 gaseous medium after stellar evolution.  This process is evoked by 
 Reisenegger et al.  (1996) to account for the metallicity of the 
 Centaurus cluster.  Their explanations perfectly account for the high 
metallicity in the center of \a85 (in the cD).  In their paper the deposit 
results from the transport from the multiphase medium to a cooler phase, 
 and the ejected gas heats the surrounding medium. This may account for 
 the central temperature (2.8~keV), which is actually smaller than the mean 
 temperature but is not very cold (see section~\ref{GV}) .  The existence of 
 several phases presently seems to be confirmed observationally at 
 least in the Virgo (Lieu et al. 1996a) and Coma 
 (Lieu et al. 1996b) clusters.

Besides, detectors such as the IPC and MPC of the Einstein 
Observatory, or the PSPC of 
ROSAT have different energy windows.  If we consider the spectral analysis, 
results undertaken with these detectors on the X-ray gas of \a85 give 
temperatures of $\sim$9~keV with the IPC, $\sim$6~keV for the MPC, and  
$\sim$4~keV for the PSPC (as indicated in section~\ref{GV}).
We can wonder if there are not gaseous phases 
with different temperatures (or a range of temperatures) that are detected
by one or another of these instruments in a privileged way. However, the 
correspondence noticed above in section~\ref{GV} between a temperature of
$\sim$4~keV and the 
velocity dispersion (760~km/s) estimated for this cluster suggests that
this temperature is correct, or at least is the temperature of the
dominant phase.  
\subsection{ Comments on $\Omega$}
The baryonic density of the Universe is obtained from the observation 
of the primordial number ratio of deuterium to hydrogen. 
Recent controversies (Rugers \& Hogan 1996, Schramm \& Turner 1996, Tytler et 
al. 1996) lead to two incompatible values of $\Omega_{b}$~:
$$\Omega_{b}=0.028 h_{50}^{-2}~~~{\rm or}~~~\Omega_{b}=0.1 h_{50}^{-2}$$
Suppose for a while that clusters -- in particular \a85 -- are a fair 
representation of the Universe. The density ratio of the baryonic 
component to the dynamical one in the cluster would then be equal to 
$\Omega_{b}$.  Comparing these values to the 17$\%$ of baryonic mass 
in the cluster leads to~: 
\begin{equation}
\Omega \simeq 0.16h_{50}^{-.5}~~~{\rm or}~~~\Omega \simeq 0.6h_{50}^{-.5}
	\label{omega}
\end{equation} 
\begin{figure}
\centerline{\psfig{figure=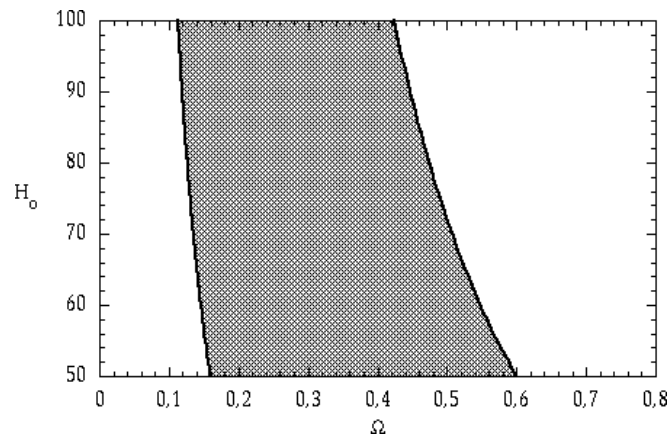,height=6cm,clip=on}}
\caption[ ]{$\Omega$ as function of $H_{0}$ deduced from the ratio of baryonic 
matter to dynamic matter at R$_{\rm L}$ (see text). The two curves correspond 
to the two values used for $\Omega_{b}$.  }
\protect\label{crisis}
\end{figure}
These values are quite similar to those given by David et al. (1995)
The high value we found compared to their high limiting value is due to 
the `new' value for $\Omega_{b}$ quoted above. We indicate in 
Fig.~\ref{crisis} 
the region of the $H_{0}$ and $\Omega$ parameters defined by 
equation~\ref{omega}. Although the amount of baryons is high in 
\a85, it is nevertheless compatible with a dense Universe. 
Furthermore, clusters, and in particular \a85, are certainly not fair 
representations of the mean Universe; for instance we have found in \a85 
a galaxy luminosity function much steeper than in the field (Durret et al. 
1996). But unless we favor an $\Omega = 1$ Universe, the `baryonic 
crisis' (White et al. 1993) does not seem dramatic. 

\subsection{Conclusions}\label{conclusions}
We have shown in this paper that, by combining the use of an X-ray telescope
with good spatial resolution, such as the PSPC on ROSAT, with a sophisticated 
imaging analysis technique such as wavelets, we have gone one step further in 
our understanding of \a85. The rather simple image we had of this X-ray 
cluster, considered until recently as a main structure with a 
smaller one south of it, is now obsolete.

The main part of the X-ray emission is due to the 
contribution of two clusters very distant from each other, almost on the 
same line of sight.  The emission of the cluster ABCG~85 itself
appears as coming from a relaxed cluster, with a weak ellipticity.  
 Our representation therefore strongly contradicts 
that presented by Kneer et al. (1996), who assume the fusion of a group (the 
south blob) with the main cluster, since we have shown that the 
south blob is resolved into very small features; the principal of 
these features is due to an over--luminosity originating in the X-ray emitting 
galaxy noted BCM$_{2}$, which forms a pair with the cD.

The detailed optical study of \a85 undertaken by acquiring both photometry and 
a large number of redshifts will enable us to discuss in more detail
the existence of substructure in \a85 and the kinematical and dynamical 
properties of this cluster (Durret et al. in preparation).  

It is apparent from our results that a good understanding of clusters
requires both good X-ray data and extensive redshift catalogues. Obviously,
new satellites such as XMM or AXAF will be of great help to 
model in detail the physical properties of clusters in a near future. However,
this analysis of \a85 shows that by combining a good morphological analysis
of ROSAT PSPC images performed with wavelets and a modelling of the X-ray
gas such as that presented here, it is already possible to change 
dramatically our view of this cluster. We therefore intend to apply these 
techniques to a sample of clusters observed with the ROSAT PSPC in a near 
future.

\begin{acknowledgements}
We acknowledge helpful e-mail exchanges with S.~Snowden concerning the use of 
his software. We thank J.-M.~Alimi, J.-P.~Chi\`eze, B.~Guiderdoni, 
N.~Prantzos, R.~Tessier and A.~Vikhlinin for enlightening discussions. We are 
very grateful to the referee, M.~Pierre, for many
interesting suggestions which helped to improve the paper. Finally, we thank
G.~Steigman for pointing out a mistake.\\
G.B.L.N. thanks Alexander v.~Humboldt Stiftung.
We acknowledge financial help from GDR Cosmologie, CNRS.
\end{acknowledgements}


\begin{thebibliography}{}

\bibitem {} Abell G.O. 1958, ApJS 3, 211
\bibitem {} Bersanelli M., Bouchet F.R., Efstathiou G. et al. 1996, ESA
Report D/SCI(96)3
\bibitem {} Biviano A., Durret F., Gerbal D. et al. 1996, A\&A 311, 95
\bibitem {} Blanchard A., Mamon G.A., Valls--Gabaud D., 1992, A\&A 264, 365
\bibitem {} Briel U.G., Henry J.P.  1994, Nature 372, 439  
\bibitem {} Cowie L.L., Binney J. 1977, ApJ 215, 723
\bibitem {} David L.P., Jones C., Forman W. 1995, ApJ 445, 578
\bibitem {} David L.P., Slyz A., Jones C., Forman W., Vrtilek S.D. 1993, ApJ 
412, 479
\bibitem {} Dickey J.M., Lockman F.J. 1990, ARA\&A 28, 215
\bibitem {} Durret F., Gerbal D., Lachi\`eze-Rey M., Lima-Neto G., Sadat R.
1994, A\&A 287, 733
\bibitem {} Durret F., Felenbok P., Gerbal D. et al., 1996, Proc. ESO
Conference ``The Early Universe with the VLT'' 
\bibitem {} Edge A.C., Stewart G.C. 1991, MNRAS 252, 428 
\bibitem {} Edge A.C., Stewart G.C., Fabian A.C. 1992, MNRAS 258, 177
\bibitem {} Fabian A.C.,  Nulsen P.E.J. 1977, MNRAS 180, 479
\bibitem {} Fabian A.C.,  Nulsen P.E.J., Canizares C.R. 1984, ARA\&A 310, 733
\bibitem {} Fukazawa Y., Ohashi T., Fabian A.C. 1994, PASJ 46, L55
\bibitem {} Forman W., Jones C. 1982, ARA\&A 20, 547 
\bibitem {} Gerbal D., Durret F., Lima-Neto G., and Lachi\`eze-Rey M. 
1992, A\&A 253, 77
\bibitem {} Godon P., Soker N., White III R.E., Regev O. 1994, AJ 108, 2009 
\bibitem {} Henriksen M.J., White, R.E. III 1996, ApJ 465, 515
\bibitem {} Henry J.P., Briel U.G.  1995, ApJ 443, L9
\bibitem {} Hirayama Y., Tanaka Y., Kogure T. 1978, Prog. Theor. Phys. 59, 751
\bibitem {} Jones C., Forman W. 1984, ApJ 276, 38
\bibitem {} Kneer R., B\"ohringer H., Neumann D., Krautter J.  1996, 
MPE Report 263, 593
\bibitem {} Koyama K., Takano S., Tawara Y. 1991, Nature 350, 135
\bibitem {} Leir A.A., van den Bergh S. 1977, ApJS 34, 381
\bibitem {} Liang H.  1995, PhD Thesis, Australian National University 
\bibitem {} Lieu R., Mittaz J.P.D., Bowyer S., et al. 1996a, ApJ 458, L5
\bibitem {} Lieu R., Mittaz J.P.D., Bowyer S., et al. 1996b, Preprint
\bibitem {} Matilsky T., Forman W., Jones C. 1985, ApJ 291, 621
\bibitem {} Mellier Y., Mathez G. 1987, A\&A 175, 1
\bibitem {} Mewe R., Lemen J.R., van den Oord G.H.J. 1986, A\&AS 65, 511
\bibitem {} Mitchell R.J., Ives J.C., Culhane J.L. 1977, MNRAS 181, 25P 
\bibitem {} Morrison R., McCammon D. 1983, ApJ, 270, 119 
\bibitem {} Prestwich A.H., Daines S. 1994, BAAS 185, 74.02
\bibitem {} Prestwich A.H., Guimond S.J., Luginbuhl C.B., Joy M.  1995, 
ApJ 438, L71
\bibitem {} Raymond J.C., Smith B.W. 1977, ApJS, 35, 419
\bibitem {} Reisenegger A., Miralda-Escud\'e J., Waxman E. 1996, ApJ 457, L11
\bibitem{} Ru\'e F., Bijaoui A. 1996, {\em Experimental Astronomy} submitted
\bibitem{} Rugers M., Hogan C.J. 1996, ApJ 459, L1
\bibitem{} Schindler S., Hattori M., Neumann D.M., B\"ohringer H. 1996,
A\&A in press
\bibitem{} Schindler S., Wambsganss J. 1996, A\&A, 313, 113 
\bibitem{} Schramm D.N., Turner M.S. 1996, Nature 381, 193
\bibitem {} Serna A., Gerbal D. 1996, A\&A 309, 65
\bibitem {} Slezak E., Durret F., Gerbal D. 1994, AJ 108, 1996
\bibitem {} Snowden S.L., McCammon D., Burrows D.N., Mendenhall 
J.A.  1994, ApJ, 424, 714
\bibitem {} Stewart G.C., Fabian A.C., Jones C., Forman W. 1984, ApJ 285, 1
\bibitem {} Struble M.F., Rood A.J. 1987, ApJ 271, 422
\bibitem{} Tytler D., Fan X.-M., Burles S. 1996, Nature 381, 207
\bibitem{} White S.D.M., Navarro J.F., Evrard A.E., Frenk C.S. 1993, Nature
366, 429
\bibitem {} Yahil A., Ostriker J.P. 1973, ApJ 185, 787
\bibitem {} Zimmerman H.U., Becker W., Belloni T. et al. 1994, EXSAS Users'
Guide, MPE Report 244
\end{thebibliography}
\end{document}